\documentclass[preprint]{elsarticle}

\usepackage{graphicx}
\usepackage{amssymb}

\journal{Journal of Magnetism and Magnetic Materials}

\begin{document}

\begin{frontmatter}

\title{A phenomenological model for the spontaneous exchange bias effect}

\author[AA]{L. Bufai\c{c}al\corref{Bufaical}}, \author[BB]{L. T. Coutrim}, \author[CC]{E. M. Bittar}, \author[CC]{F. Garcia}, \ead{lbufaical@ufg.br}

\address[AA]{Instituto de F\'{i}sica, Universidade Federal de Goi\'{a}s, 74001-970 , Goi\^{a}nia, GO, Brazil}
\address[BB]{Departamento de F\'{\i}sica, Universidade Federal do Rio Grande do Norte, 59078-900, Natal, RN, Brazil}
\address[CC]{Centro Brasileiro de Pesquisas F\'{\i}sicas, Rua Dr. Xavier Sigaud 150, 22290-180, Rio de Janeiro, RJ, Brazil}

\cortext[Bufaical]{Corresponding author}

\begin{abstract}

In this work we propose an alternative model to explain the spontaneous exchange bias (SEB) effect observed in spin glass (SG)-like systems. As in a previously proposed model \cite{ZEBmodel}, it is based on the unconventional dynamics of the SG-like moments at the magnetic hysteresis cycle. However, using a reliable estimate of the amount of SG-spins that are relaxing during the cycle, the new model can correctly describe the changes in the SEB observed for measurements performed at different temperatures and different maximum applied fields.

\end{abstract}


\begin{keyword}
Exchange Bias; Spin-glass; Double-perovskite
\end{keyword}

\end{frontmatter}

\section{INTRODUCTION}

The exchange bias (EB) effect finds its applicability in magnetic recording read heads and spintronic devices. The phenomena is known since the 1950's, being characterized by a horizontal shift of the magnetic hysteresis loop of heterostructured materials \cite{Nogues2}. In general, the exchange unidirectional anisotropy is set at the interface of different magnetic phases after the system is cooled in the presence of an external magnetic field ($H$) from above its magnetic transition temperature ($T$). Nonetheless, recently there were discovered some materials manifesting the EB spontaneously, \textit{i.e.} even after being cooled from an unmagnetized state down to low $T$ in zero $H$ \cite{Wang,Maity,CoIr_PRB}.

Recently we have shown that the presence of a re-entrant spin glass (RSG) state is necessary for the manifestation of the spontaneous EB (SEB) effect, and proposed a model to explain the phenomenon \cite{ZEBmodel}. The model is based on the pinning of the spin glass (SG)-like moments and on their unusual temporal evolution in magnetization as a function of $H$ [$M(H)$] curves, from which it is obtained accurate predictions of the SEB effect observed in two representative SEB materials, La$_{1.5}$Sr$_{0.5}$CoMnO$_{6}$ \cite{Murthy} (LSCMO) and La$_{1.5}$Ca$_{0.5}$CoMnO$_{6}$ \cite{CaCoMn_JMMM,PRB2019} (LCCMO). More specifically, our model considers an asymmetric magnetic relaxation of the SG-like moments at the regions encompassing the positive and negative coercive fields ($H_C$) in the $M(H)$ cycle of these materials. It is assumed, as an approximation, that for the region close to the positive $H_C$, half of SG-like spins are relaxing due to the negative $H$ previously applied in the third quadrant of the $M(H)$ cycle, while the other half is pinned toward the positive $H$ direction. However, in spite of the very good agreement between the theoretical and experimental results observed for LSCMO and LCCMO in $M(H)$ loops carried at $T$ = 5 K with a maximum applied field ($H_m$)  of 90 kOe, one cannot rely that this assumption is always correct. By measuring at different $T$ and different $H_m$, the correlation lengths for the SG-like phase can change, resulting in unique magnetic relaxations and consequently in distinctive EB fields ($H_{EB}$).

Due to its robust SEB effect \cite{ZEBmodel,Murthy}, in this work we chose the LSCMO compound as a representative example of a SEB material to confirm that $T$ and $H_m$ remarkably affect the $H_{EB}$ observed in $M(H)$ loops, complementing the model proposed in Ref. \cite{ZEBmodel}. Here we offer a complementary model, adapted from the previous one, but that is based on a plausible estimate of the SG-like moments pinned toward the positive and negative $H$ directions. This model can capture the evolution of $H_{EB}$ as a function of $T$ and $H_{m}$, resulting in theoretical values that are very close to those experimentally observed for LSCMO.

\section{EXPERIMENT DETAILS}

The polycrystalline LSCMO sample here investigated was prepared as described elsewhere \cite{ZEBmodel}. The $M(H)$ loops were carried out at several $T$ and $H_m$ using a Quantum Design PPMS-VSM magnetometer. The curves were measured at a $H$ sweep rate of 210 Oe/s, after zero field cooling (ZFC) the system. In order to prevent the presence of trapped current on the magnet and ensure a reliable ZFC process, from one measurement to another the sample was warmed up to the paramagnetic state and the coil was demagnetized in the oscillating mode.

\section{RESULTS AND DISCUSSION}

\begin{figure}
\begin{center}
\includegraphics[width=0.8 \textwidth]{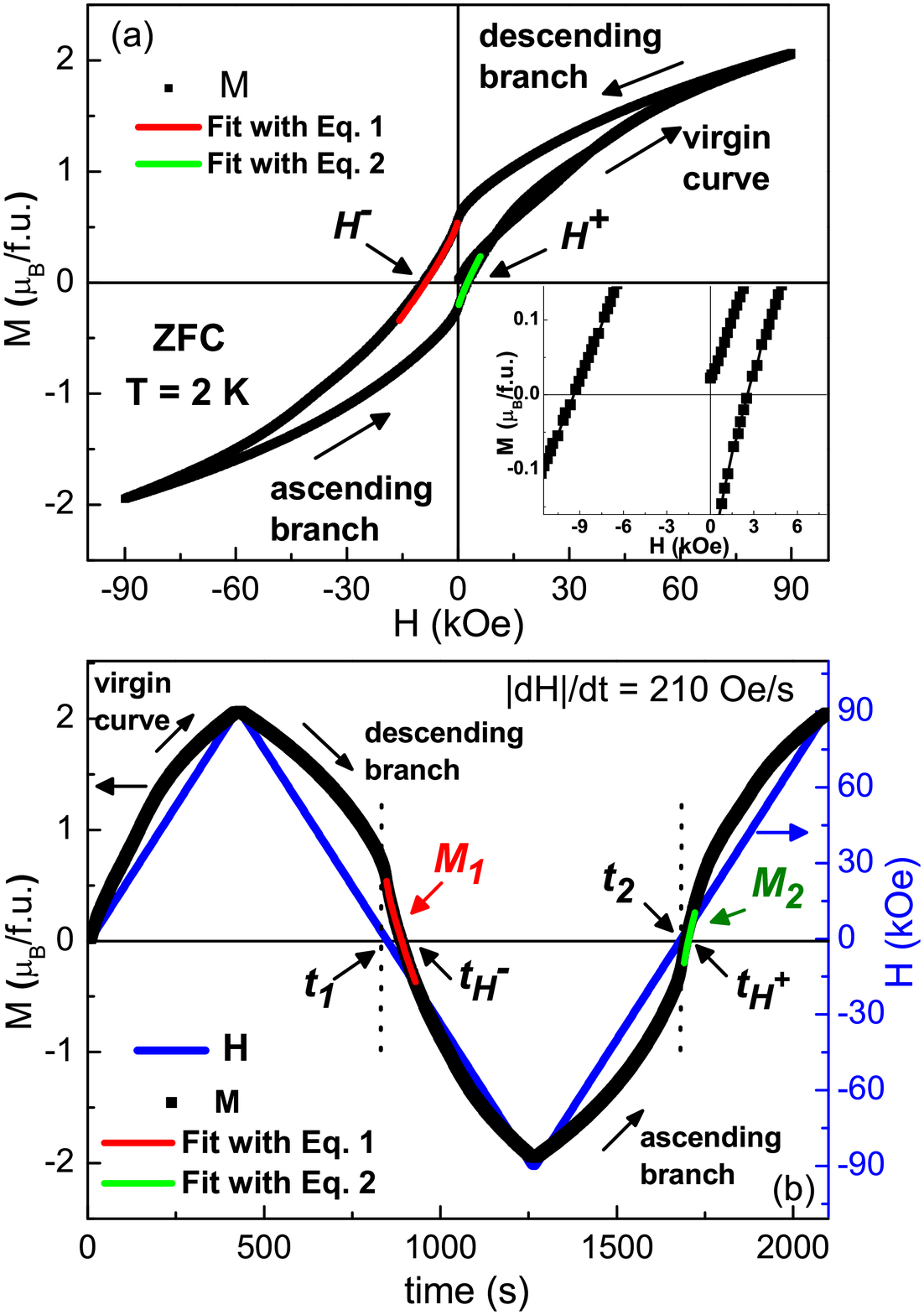}
\end{center}
\caption{(a) $M(H)$ loop of LSCMO at $T=2$ K and $H_m$ = 90 kOe. Red and green solid lines are the calculated $M_{1}$ (Eq. \ref{Eq1}) and $M_{2}$ (Eq. \ref{Eq2}) stretches, respectively. Inset shows zoom in around $M=0$, evidencing the shift along $H$-axis. (b) The same hysteresis loop for LSCMO, now displayed in $M(t)$ mode. The blue solid line is the magnetic field as function of time.}
\label{Fig_MxH}
\end{figure}

Fig. \ref{Fig_MxH}(a) shows the $M(H)$ loop at $T$ = 2 K and $H_{m}$ = 90 kOe. The shape of the curve is a result of the contribution of three distinct magnetic phases, a ferromagnetic (FM), an antiferromagnetic (AFM) and a SG-like phase \cite{ZEBmodel}, producing a closed loop that is asymmetric in respect to the $H$ axis, as evidenced in the inset. In the $M(H)$ cycle the magnetization ($M$) depends on $H$, which in turn varies linearly with time ($t$). Thus, the hysteresis curves can be displayed in the form of $M$ as a function of $t$ [$M(t)$], Fig. \ref{Fig_MxH}(b). Since our model is based on the time-evolution of the magnetization of the SG-like phase, this form is suitable for its understanding.

The phenomenological model intends to describe the $H_{EB}$ obtained from $M(H)$ curves, herein defined as $H_{EB}=|H^{+}+H^{-}|/2$, where $H^{+}$ and $H^{-}$ are the positive and negative coercive fields, respectively. For that, the $M_1$ and $M_2$ stretches of the curve, encompassing respectively the $H^{-}$ and $H^{+}$ [see Figs. \ref{Fig_MxH}(a) and (b)], must be calculated. The material's net magnetization results from the contributions of FM, AFM and SG-like phases, leading to the following equation for $M_1$
\begin{equation}
M_{1}(t) = \{M_{sp} + M_{0}e^{-\left[(t-t_{1})/t_{p}\right]^{n}}\} - \{A(t-t_{1}) + B(t-t_{1})^{r}\}, \label{Eq1}
\end{equation}
which corresponds precisely to the equation proposed in Ref. \cite{ZEBmodel}. The first pair of braces represents the SG-like phase's relaxation from the previously applied positive $H_m$. In fact, this corresponds to the stretched exponential equation commonly used to verify the isothermal remnant magnetization ($IRM$) of glassy magnetic systems \cite{Chamberlin}, where $M_{sp}$ represents the spontaneous magnetization of the FM phase, $M_0$ is the initial magnetization of the SG-like phase at the instant $t_1$ when $H$ = 0 (see Fig. \ref{Fig_MxH}), and $t_p$ and $n$ are the time and the time-stretch exponential constants, respectively. The second pair of braces account for the contributions of the AFM and FM phases to $M_1$, when under the effect of the immediately applied negative $H$. The $A$ parameter is related to the linear dependence of the AFM phase with $H$ (and consequently with $t$), while the $B$ and $r$ parameters account for the non-linear contribution of the FM phase to the magnetization.

To get the parameters of the first pair of braces of Eq. \ref{Eq1}, the $IRM(t)$ curve must be fitted immediately after $H_m$ is turned on and subsequently turned off in the ZFC sample. Subsequently, by fixing the parameters of the SG-like phase, the $M_1$ stretch can be fitted with Eq. \ref{Eq1}, yielding a very good match, as expected \cite{ZEBmodel}. The main results obtained from the fittings are displayed in Table \ref{T1}.

For the $M_2$ stretch, we take into account the unusually slow relaxation of glassy magnetic systems. It is assumed that at $M_2$ not all spins of the SG-like phase are relaxing due to the negative field previously applied during the $M(H)$ cycle, but it considers that some amount of them are still pinned toward positive direction due to the firstly applied positive $H_m$. The resulting equation for $M_2$ becomes
\begin{equation}
M_{2}(t) = -\{M_{sp} + xM_{0}e^{-\left[(t-t_{2})/t_{p}\right]^{n}}\}+\{(1-x)M_{0}e^{-\left[(t-t_{1})/t_{p}\right]^{n}}\} + \{A(t-t_{2}) + B(t-t_{2})^{r}\}, \label{Eq2}
\end{equation}
where the first pair of braces represents the decay of the SG-like spins that are relaxing from the negative $H$ previously applied, the second pair corresponds to the relaxation from the positive $H$ applied before, and the third pair represents the contributions of the AFM and FM phases that are under the effect of the just applied positive $H$. This is similar to the equation proposed in Ref. \cite{ZEBmodel}, where the decay of the SG moments pointing toward the positive direction starts at $t_1$ while the decay of those pointing toward the negative direction starts at $t_2$. However, there is a remarkable difference here. While for the previous model it was considered that an equal amount of the moments were relaxing in opposite directions, here the amount of SG-spins pointing toward negative ($x$) and positive (1-$x$) directions can be estimated directly from the experimental curve. To fit $M_2$ with Eq. \ref{Eq2}, all parameters, with the exception of $x$, are kept fixed at the values obtained from the fit of $M_1$.

\begin{figure}
\begin{center}
\includegraphics[width=0.8 \textwidth]{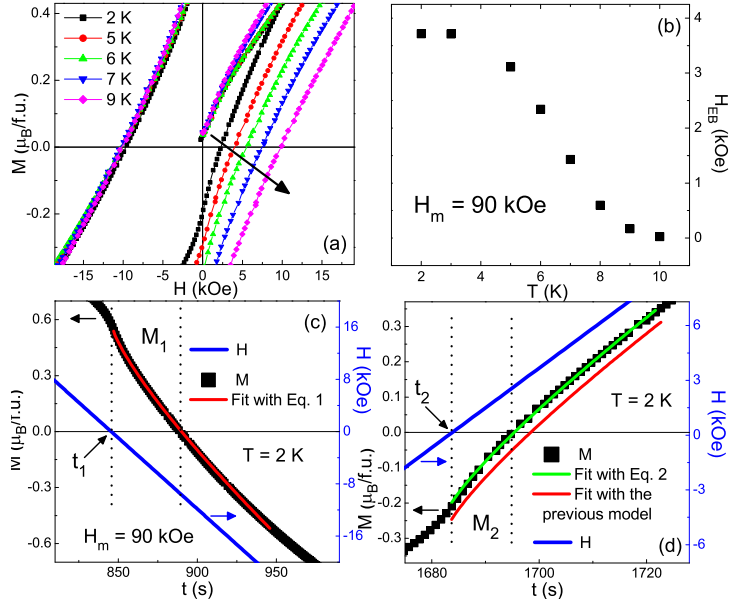}
\end{center}
\caption{(a) Magnified view of the $M(H)$ loops of LSCMO carried at $H_m$ = 90 kOe at different temperatures. (b) $H_{EB}$ as a function of $T$. (c) and (d) show magnified views of the experimental and calculated $M_1$ and $M_2$ stretches of the loop carried at $T$ = 2 K. The blue solid line represents the $H$ time dependence.}
\label{Fig_T}
\end{figure}

Although the previous model succeed in predicting the $H_{EB}$ values of $M(H)$ loops carried at $T$ = 5 K and $H_m$ = 90 kOe, changes in the $T$ and/or the $H_m$ at which the experiments are carried can lead to deviations from the calculated results in relation to the experimentally observed values. Fig. \ref{Fig_T}(a) shows a magnified view of loops carried with $H_m$ = 90 kOe at different temperatures. It is evident that changing $T$ has a greater impact on $H^{+}$, while $H^{-}$ keeps nearly unchanged. This can be understood in terms of the gain of thermal energy with increasing $T$, which favors the flipping of spins toward the negative field direction at the third quadrant of the cycle. As $T$ increases the amount $x$ of SG-spins flipped enhances, leading to the increase of $H^{+}$ and consequently to the decrease of $H_{EB}$, as can be seen in Fig. \ref{Fig_T}(b). Since the previous model assumes a fixed number of SG-spins in each direction, it cannot capture such changes. Fig. \ref{Fig_T}(d) shows the fittings of the $M_2$ stretch of the 2 K $M(H)$ curve for both the previous and the alternative model here proposed, and Table \ref{T1} displays the $H_{EB}$ obtained from each model. The difference is remarkable.

\begin{table}
\caption{Main results obtained from the fits of $M_1$ and $M_2$ stretches with Eqs. \ref{Eq1} and \ref{Eq2}. The $H_{EB}$ results obtained experimentally, the values calculated with Eqs. \ref{Eq1} and \ref{Eq2} and those obtained from the previous model of Ref. \cite{ZEBmodel} are referred respectively as $H_{EB}^{exp}$, $H_{EB}^{new}$ and $H_{EB}^{old}$.}
\label{T1}
\begin{tabular}{c|ccc}
\hline \hline
$H_m$ (kOe) & 70 & 90 & 90  \\
$T$ (K) & 2 & 2 & 5 \\
\hline
$M_{sp}$ ($\mu_B$/f.u.) & 0.220 & 0.226 & 0.231 \\
$M_{0}$ ($\mu_B$/f.u.) & 0.348 & 0.363 & 0.370\\
$t_p$ (s) & 4.882$\times$10$^{10}$ & 1.983$\times$10$^{9}$ & 1.961$\times$10$^{10}$  \\
$n$ & 0.130 & 0.122 & 0.143 \\
\hline
$A$ ($\mu_B$/f.u.) & 5.2$\times$10$^{-4}$ & 6.1$\times$10$^{-4}$ & 9.8$\times$10$^{-4}$ \\
$B$ ($\mu_B$/f.u.) & 0.028 & 0.026 & 0.023 \\
$r$ & 0.78 & 0.79 & 0.74 \\
$x$ & 0.47 & 0.43 & 0.54 \\
\hline
$H_{EB}^{exp}$ (Oe) & 3517 & 3716 & 3114 \\
$H_{EB}^{new}$ (Oe) & 3539 & 3719 & 3084 \\
$H_{EB}^{old}$ (Oe) & 3049 & 3085 & 3172 \\
\hline \hline
\end{tabular}
\end{table}

\begin{figure}
\begin{center}
\includegraphics[width=0.8 \textwidth]{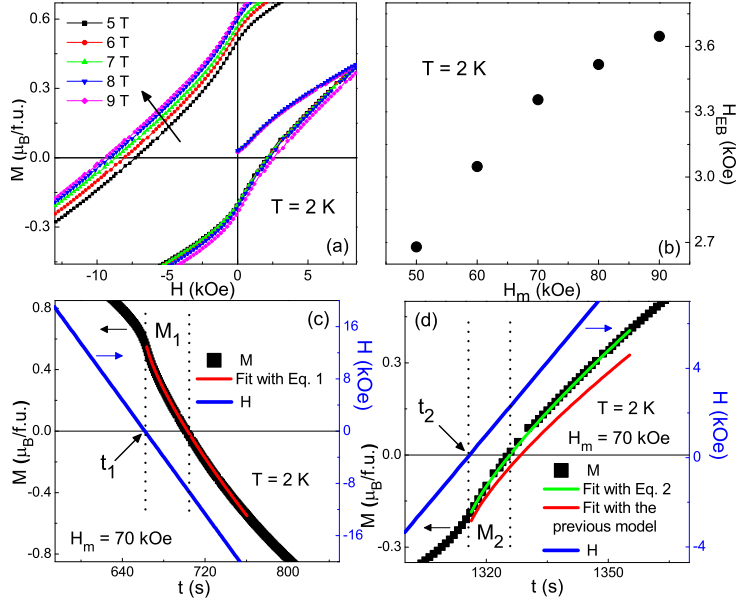}
\end{center}
\caption{(a) Magnified view of the $M(H)$ loops of LSCMO carried at $T$ = 2 K with different $H_m$. (b) $H_{EB}$ as a function of $H_m$. (c) and (d) show magnified views of the experimental and calculated $M_1$ and $M_2$ stretches of the loop carried at $H_m$ = 70 kOe. The blue solid line represents the $H$ time dependence.}
\label{Fig_Hm}
\end{figure}

The alternative model can also capture changes in $M(H)$ curves measured at different $H_m$. Fig. \ref{Fig_Hm}(a) shows that $H^{-}$ is significantly influenced by changes in $H_m$, while $H^{+}$ keeps nearly unchanged. In this case, as larger is $H_m$ greater will be the amount of spins pinned toward positive direction. This is translated in the equations by the decrease of $x$ and by the increase of both $M_{sp}$ and $M_0$, as can be seen in Table \ref{T1}. This leads to the increase of $H^{-}$, resulting in the enhancement of $H_{EB}$. Again, the previous model cannot account for such changes because it considers a fixed number of spins pointed toward opposite directions.

Despite the very good adequacy of the alternative model proposed here, one must stress that it results from approximations and simplifications, most of them concerning the dynamics of the SG-like phase. For instance, the varying $H$ may alter the balance between the distinct magnetic phases present in the system, in a way that in principle several parameters of the equations should be functions of $t$. To minimize this effect, we fitted $M_1$ and $M_2$ for short time-intervals. Another clear approximation comes from the fact that the second brace of Eq. \ref{Eq1} and the third brace of Eq. \ref{Eq2} take into account the influence of the immediately applied $H$ only on the FM and AFM phases, but not on the SG-like one. In principle, a term should be added in the equations to regard it. However, since the two terms in these braces were enough to yield precise fits of $M_1$ and $M_2$, we have disregarded a third term in order to make the model as simple as possible. Despite these simplifications, Eqs. \ref{Eq1} and \ref{Eq2} can successfully reproduce the $H_{EB}$ observed in LSCMO. Since the RSG state is a common feature of all known SEB materials, in principle the model could be applied to any similar compound. Having stablished that the glassy magnetism is imperative for the manifestation of SEB, the model here proposed can guide the search for new materials presenting robust SEB effect at higher $T$.

\section{CONCLUSIONS}

In summary, we proposed an alternative phenomenological model to explain the SEB effect observed in RSG systems. It is based on the pinning and on the dynamics of the SG-like moments relaxing during a magnetic hysteresis measurement. Differently of the model previously proposed \cite{ZEBmodel}, here we considered a reliable estimate of the amount of SG-like moments pinned toward the positive and negative $H$ directions during the $M(H)$ cycle. We used LSCMO as a representative example of SEB to check the model and show that it allows an accurate calculation of the $H_{EB}$ and a correct description of the changes observed in the SEB effect for $M(H)$ curves measured at different $T$ and $H_m$.

\section{ACKNOWLEDGMENTS}
This work was supported by Conselho Nacional de Desenvolvimento Cient\'{i}fico e Tecnol\'{o}gico (CNPq) [No. 400134/2016-0], Funda\c{c}\~{a}o Carlos Chagas Filho de Amparo \`{a} Pesquisa do Estado do Rio de Janeiro (FAPERJ), Funda\c{c}\~{a}o de Amparo \`{a} Pesquisa do Estado de Goi\'{a}s (FAPEG) and Coordena\c{c}\~{a}o de Aperfei\c{c}oamento de Pessoal de N\'{i}vel Superior (CAPES).

\end{document}